\documentclass[11pt]{article}

\newfont{\tenmsb}{msbm10 scaled\magstep1}
   
\makeatletter
\@addtoreset{equation}{section}
   
\makeatother

\newcommand\half{{\scriptstyle{\frac{1}{2}}}}

\newcommand{\p}{{\partial}}
\newcommand{\vx}{{\vec x}}

\newcommand{\vR}{{\vec R}}
\newcommand{\vQ}{{\vec Q}}

\newcommand{\vp}{{\vec p}}
\newcommand{\vb}{{\vec b}}

\newcommand{\vA}{{\vec A}}
\newcommand{\vE}{{\vec E}}

\begin{document}

\setlength{\baselineskip}{16pt}

\title{
Noncommuting coordinates in the  Hall effect
and in vortex dynamics\footnote{Talk given at the joint 
COSLAB-VORTEX-BEC2000+ Workshop. Bilbao, july'03.
}}

\author{
P.~A.~Horv\'athy
\\
Laboratoire de Math\'ematiques et de Physique Th\'eorique\\
Universit\'e de Tours\\
Parc de Grandmont\\
F-37 200 TOURS (France)
}

\date{\today}

\maketitle

\begin{abstract}
    Laughlin's Ansatz to explain the fractional Quantum Hall effect
    is derived by coupling a particle associated with ``exotic'' the two-fold 
    central extension of the planar Galilei group. The reduced system
    is identical to the one used to describe the dynamics of vortices
    in an incompressible planar fluid.
\end{abstract}


\goodbreak

\section{Introduction}

One of the most hotly debated issues of present-day
theoretical high-energy physics is noncommutative
(quantum) mechanics \cite{NCQM}, where the coordinates satisfy the
nontrivial commutation relation
\begin{equation}
    \big[\hat{x},\hat{y}\big]=i\hbar\theta.
\label{NCrel}
\end{equation}
The real number $\theta$ here is referred to as the
noncommutative parameter.
Such a relation may appear rather puzzling at the first sight,
and one can wonder about the physical motivations.
High-energy physicist
usually refer to higher dimensional branes and strings;
this can however leave ordinary physicsts somewhat sceptical.
Below we present some arguments in favor of
(\ref{NCrel}) which are, hopefully, more convincing for 
down-to-earth physicsists.

\section{The Fractional Quantum Hall Effect}

The  main experimantal result about the Fractional Quantum Hall Effect 
(FQHE) is
that the Hall and diagonal resistivity of some heterostructures is
\begin{equation}
    R_{xy}=\frac{1}{\nu}\frac{h}{e^2},
    \qquad
    R_{xx}=0
\label{fracquant}
\end{equation}
where the filling factor $\nu$ is an odd integer, 
$\nu=2n-1$ \cite{Ezawa, QHE}.
In his seminal paper Laughlin \cite{LAUGH} argues that the FQHE can entirely 
be explained within the lowest Landau level~: the system condensates 
into a collective ground state, representing an incompressible
quantum fluid, based on the ``Laughlin'' wave functions
\begin{equation}
    \psi(z)=f(z)e^{-B|z|^2/4}
    \label{Lwavefunc}
\end{equation}
where $f(z)$ is analytic. The fractional quantization conditions
(\ref{fracquant}) is recovered when $f(z)=z^{2n+1}$.

This Note, based on joint work with
Christian Duval (and Zal\'an Horv\'ath) \cite{DH, DHH}, 
aims justify the starting point of Laughlin's description 
from first principles. Our work ends where that of Laughlin begins. 

Before presenting our theory, let us recall the usual treatment of
the Landau problem \cite{Ezawa}. A charge confined to the plane and 
moving under the influence of a perpendicular magnetic and a planar
electric field is described by the Hamiltonian $H=\vp^2/2m+eV(\vx)$
where $[p_{1},p_{2}]=i\hbar eB$. When $V=0$, the spectrum is
$E_{n}=\hbar (eB/m)(\half+n\big)$.

Classically, the particle
performs helical motion, as seen from the decomposition
\begin{equation}
    \vQ=\vx-\vR,\qquad
    R_{i}=\frac{1}{eB}\varepsilon_{ij}p_{j}.
    \label{guidcentcoord}
\end{equation}
In fact, 
$\vQ$ follows the Hall law, and $\vR$ performs a uniform
rotation.
The remarkable fact \cite{Ezawa} is that the guiding center coordinates
do not commute but satisfy rather
\begin{equation}
    \big[\hat{Q}_{1},\hat{Q}_{2}\big]=-i\frac{\hbar}{eB}
\end{equation}
that realize the commutation relation (\ref{NCrel}) with
$\theta=-(eB)^{-1}$.

The Landau spectrum is explained by the decomposition (\ref{guidcentcoord}):
the guiding center contributes the ground state energy, and the higher
Landau levels come from the oscillations of the internal coordinate
$\vR$. Semiclassically, $<\vR^2>=(1+2n)\frac{\hbar}{eB}$ \cite{Ezawa}.

It is worth noting that, for {\it very special initial conditions},
the guiding center motion can be materialized by actual motions.
If the initial postion and velocity are such that the electric field
is compensated by the Lorentz force, $eE_{i}+eB\varepsilon_{ij}v_{j}=0$,
then the motion is in fact at right angle to $\vE$ i. e., along an 
equipotential.
The generic motion is, of course, the helical one; the initial 
conditions which satisfy the force-free conditions 
form indeed a two-dimensional surface in $4D$ phase space.

Intuitively, our theory presented below has the peculiarity to eliminite these
generic, helical motions,  leaving us only with those of the guiding center.

\section{Exotic particles}\label{Exotic}

Let us now present our model. Following Wigner \cite{Wigner}, 
elementary particles
correspond to irreducible representations of their fundamental
symmetry groups. In the nonrelativistic case, however, 
the Galilei group is only represented projectively, i.~e., 
only up-to-phase~: in spatial dimensions
at least $3$, it is only a one-parameter
central extension of the Galilei group that is unitarily 
represented. It has been shown furthermore by Bargmann \cite{Bargmann} 
that this phase can not be elimininated by any redefinition, as it
corresponds to a nontrivial cohomology class of the group, labeled
by the real parameter $m$, interpreted as the mass.
Let us record for further reference that a Galilean
boost with parameter $\vb$ is implemented, 
in the momentum representation, as
\begin{equation}
    U_{\vb}\phi(\vp)=\phi(\vp-m\vb).
    \label{ordboost}
\end{equation}
It follows that the components of the boost generator,
$\widehat{g}_{i}=mi\p_{p_{i}}$ commute, 
$\big[\widehat{g}_{1},\widehat{g}_{2}\big]=0$.

The planar case is instead rather peculiar in that the cohomology
is {\it two dimensional} with generators $m$ and $\kappa$,
respectively \cite{exogal}.
This has been noticed a long time ago but has not been 
sufficiently appreciated until recently.

Now the geometric quantization of Kirillov-Kostant-Souriau \cite{GQ,SSD}
associates the representations of a group with the coadjoint orbits,
endowed with their canonical symplectic structure.
The idea os Souriau has been furthermore to consider these orbits
as underlying classical models. Explicit calculation 
\cite{planarorbit,DH}
yields that the orbit is $4$ dimensional, parametrized with
the position and momentum, $\vx$ and $\vp$, and carries the
``exotic'' symplectic structure
\begin{equation}
    \omega=d\vp\wedge d\vx+
    \frac{\theta}{2}\varepsilon_{ij}dp_{i}\wedge dp_{j},
    \label{exosymp}
\end{equation}
where we wrote
$
\theta={\kappa}/{m^2}.
$
Using again the momentum representation, a Galilean boost is now 
represented by 
\begin{equation}
    U_{\vb}\phi(\vp)=e^{im\theta\vb\times\vp}
    \phi(\vp-m\vb)
    \label{exoboost}
\end{equation}
cf. {ordboost}.
The inclusion of the phase factor implies that the components
of new boost generator,
\begin{equation}
\widehat{g_j}	=
	m\left[i\displaystyle\frac{\ \p }{\p p_{j}}
	+\displaystyle\frac{1}{2}\theta\varepsilon_{jk}\,p_{k}\right],
\label{exoboostgen}
\end{equation}
satisfy rather the ``exotic'' commutation relation
\begin{equation}
\big[\hat{g}_{1},\hat{g}_{2}\big]=-i\hbar m^2\theta.
\end{equation}

\goodbreak

Having constructed our free model, let us couple it minimally
to an electromagnetic field by considering the action
\begin{equation}
 \int{
(\vp-e\vA\,)\cdot d\vx
-\frac{\vp{\,}^2}{2m}+eV\,dt
+
\frac{\theta}{2}\,\vp\times d\vp
},
\label{matteraction}
\end{equation}
where $(V,\vA)$ is an electro-magnetic
potential. The associated Euler-Lagrange equations read
\begin{equation}
\begin{array}{rcl}\displaystyle
m^*\dot{x}_{i}
&=&
p_{i}-m\theta e\,\varepsilon_{ij}E_{j}
\\[8pt]
\dot{p}_{i}
&=&
eE_{i}+eB\,\varepsilon_{ij}\dot{x}_{j}
\end{array}
\label{eqmotion}
\end{equation}
where we have introduced the \textit{effective mass}
\begin{equation}
m^*=m(1-e\theta B).
\label{effmass}
\end{equation}

Let us observe that
the velocity and momentum
are not proportional if $\theta\neq0$.

The equations of motions (\ref{eqmotion}) can also be written as
\begin{equation}
	\omega_{\alpha\beta}\dot{\xi}_\beta=\frac{\p h}{\p \xi_\alpha}
\qquad\hbox{where}\qquad
	\big(\omega_{\alpha\beta}\big)=
	\left(\begin{array}{cccc}
	0&\theta&1&0\\[2mm]
	-\theta&0&0&1\\[2mm]
	-1&0&0&B\\[2mm]
	0&-1&-B&0\\[2mm]
	\end{array}\right).
\label{symplecticmatrix}
\end{equation}
Note that the electric and magnetic fields
are otherwise arbitrary solutions of the homogeneous Maxwell equation
$\p_tB+\varepsilon_{ij}\p_iE_j=0$, which guarantees that the
two-form $\omega=\half\omega_{\alpha\beta}d\xi^\alpha\wedge{}d\xi^\beta$
is closed, $d\omega=0$.
It follows that the associated Poisson bracket satisfies the Jacobi 
identity.
\goodbreak

When  $m^*\neq0$, the determinant
\begin{equation}  
\det\big(\omega_{\alpha\beta}\big)=
\left(1-\theta\, B\right)^2
=\big(\frac{m^*}{m}\big)^2
\end{equation}
is nonzero;  the matrix  $(\omega_{\alpha\beta})$ in (\ref{symplecticmatrix})
 is indeed symplectic and can therefore be inverted.
 Then  the equations of motion
(\ref{symplecticmatrix}) (or (\ref{eqmotion}))
take the Hamiltonian form 
$\dot{\xi}_{\alpha}=\big\{\xi_{\alpha}, h\big\}$,
with the standard Hamiltonian, but with the ``exotic'' Poisson
bracket $\{f,g\}=(\omega^{-1})_{\alpha\beta}\p_{\alpha}f\p_{\beta}g$.
The fundamental commutation relations are in particular 
 \begin{equation}
\begin{array}{lll}
\{x_{1},x_{2}\}=
\displaystyle\frac{m}{m^*}\,	\theta,
	\\[3.4mm]
	\{x_{i},p_{j}\}=\displaystyle\frac{m}{m^*}\,\delta_{ij},
	\\[3.4mm]
	\{p_{1},p_{2}\}=\displaystyle\frac{m}{m^*}\,B.
\end{array}
\label{Bcommrel}
\end{equation}

\goodbreak

Further insight can be gained when the magnetic field $B$
is a (positive) nonzero constant.
The vector potential can then be chosen
as $A_i=\half{}B\varepsilon_{ij}\,x_{j}$, the electric field $E_i=-\p_iV$ being
still arbitrary. Let us introduce the new coordinates
\begin{equation}
Q_{i}=x_{i}+
\displaystyle\frac{1}{eB}\Big([1-\sqrt{
\displaystyle\frac{m^*}{m}}\,\Big)
\varepsilon_{ij}\,p_{j}.
\label{goodcoordinates}
\end{equation}
Then the equations of motion (\ref{eqmotion}) are conveniently presented
in terms of the new variables $\vQ$ and the old momenta $\vp$, as
\begin{equation}
    \left\{\begin{array}{ll}
\dot{Q}_{i}=\varepsilon_{ij}\displaystyle\frac{E_{j}}{B}
+\displaystyle\sqrt{\frac{m}{m^*}}\left(
\frac{p_{i}}{ m}-\varepsilon_{ij}\displaystyle\frac{E_{j}}{B}\right),
\\[4mm]
\dot{p}_{i}=
\varepsilon_{ij}{B}\displaystyle\frac{m}{m^*}\left(
\frac{p_{j}}{ m}-
\varepsilon_{jk}\displaystyle\frac{E_{k}}{B}\right).
\end{array}\right.
\label{Qpeqmot}
\end{equation}

When the magnetic field takes the particular value
\begin{equation}
B=B_{c}=\frac{1}{e\theta},
\label{critB}
\end{equation}
the effective mass (\ref{effmass}) vanishes, $m^*=0$, so that
the system becomes singular.
Then the time derivatives $\dot{\xi_\alpha}$ can no longer be 
expressed from the
variational equations (\ref{symplecticmatrix}), and we have resort to
``Faddeev-Jackiw'' reduction \cite{FaJa}. The result is \cite{DH}
that the momentum stops to be a dynamical variable,
\begin{equation}
\frac{p_i}{m}-\varepsilon_{ij}\frac{E_j}{B_{c}}=0,
\label{pHall}
\end{equation}
and we 
 end up with the reduced Lagrangian
\begin{equation}
L_{\rm red}=\frac{1}{2\theta}\vQ\times\dot{\vQ}-eV(\vQ),
\label{redlag}
\end{equation}
supplemented with the Hall constraint (\ref{pHall}).
Thus, the $4$-dimensional phase space is reduced to $2$ dimensions, with
$Q_1$ and $Q_2$
as canonical coordinates, and reduced symplectic two-form
$ 
\omega_{\rm red}=
\half eB_{c}\,\varepsilon_{ij}dQ_{i}\wedge dQ_{j}.
$
The new coordinates are therefore again non-commuting,
\begin{equation}
\big\{Q_{1}, Q_{2}\big\}_{\rm red}=-\theta=-\frac{1}{eB_{c}}.
\label{redcommrel}
\end{equation}

Remarkably, our new coordinates become, for $m^*=0$, precisely the guiding center
coordinates (\ref{guidcentcoord}), as anticipated by the notation.

The equations of motion associated with (\ref{redlag}), and also consistent with
the Hamilton equations $\dot{Q}_i=\big\{Q_i,H\}_{\rm red}$, are given by
\begin{equation}
\dot{Q}_{i}=\varepsilon_{ij}\frac{E_{j}}{B_{c}},
\label{QHall}
\end{equation}
consistently with the Hall law.
Putting $B_{c}=1/e\theta$, the Lagrangian (\ref{redlag}) becomes formally
identical to the one derived by
Dunne et al.~\cite{DJT}  letting the {\it real} mass
go to zero. 
\goodbreak

Quantization of the reduced system is conveniently carried out in the
Bargmann-Fock representation \cite{DH}. Setting 
$z=Q_{1}+iQ_{2}$,
the (reduced) wave functions are precisely those of
Laughlin (\ref{Lwavefunc}). 
The reduced position operators are
\begin{equation}
	\widehat{z}=\,zf,
\qquad
	\widehat{\bar{z}}\,f=
	2\,\p_zf,
\label{redquantumposition}
\end{equation}
whose commutator is
$[\widehat{z},\widehat{\!\bar{z}}]=2/eB$.
Finally, the reduced Hamiltonian is just the potential
$eV(z,\bar{z})$.
In conclusion, we recover the
``Laughlin'' description \cite{QHE}
of the ground states of the FQHE.

Interestingly, similar ideas to ours have been expressed,
independently, by Fosco and Lopez \cite{FoLo}.

Let us mention that a fluid model can be built on our
``exotic mechanics'' following the general principles
of plasma physics; in the critical case, it yields
an incompressible quantum fluid that moves collectively
according to the Hall law \cite{DHH}.

\goodbreak

\section{Dynamics of planar vortices}

It has been known for over hundred years that fluid vortices in the plane
follow a simple, first-order non-newtonian dynamics \cite{Kirch, 
HLamb, Yourgrau}.
For the sake of simplicity, we restrict ourselves to two 
vortices of identical vorticity. 
The center-of-vorticity coordinates are constants of the motion. 
For the relative coordinates 
$x=x_{1}-x_{2}$ and $y=y_{1}-y_{2}$, respectively,
the equations of motion become
\begin{eqnarray}
    \gamma\,\dot{x}=\partial_{y} H,
    \qquad
    \gamma\,\dot{y}=-\partial_{x} H,
     \label{effdyn}
\end{eqnarray}
where $\gamma$ is the vorticity. Let us stress that, in the present
purely hydrodynamic context, $\gamma$ can be any real number.
The Hamiltonian, representing the interaction of the vortices, reads 
\begin{equation}  
H=-\frac{\gamma^2}{4\pi}\ln r.
\label{vortham}
\end{equation}

These equations can be derived from the hydrodynamics of an 
incompressible planar fluid \cite{NCvort, SSVW}.
The important fact for our purposes is that
Eq. (\ref{effdyn}) is a Hamiltonian system, 
\begin{equation}
\dot{\xi}=\big\{\xi,H\big\},
\qquad
\xi=(x,y),
\label{hameq}
\end{equation}
where the Poisson bracket associated with the symplectic structure
$\Omega=
\gamma\,dx\wedge dy.
$
Thus, planar vortex dynamics is exactly of the form of our ``reduced
dynamics'' presented in Section \ref{Exotic}.
This ``coincidence'' underlines the fundamental role of vortices
in explaining the FQHE.

It is worth noting that the hydrodynamic formulae above  apply to 
the effective dynamics of point-like vortices in a thin film of
superfluid ${}^4$He \cite{HMC}; the only difference being that $\gamma$,
the strength of the vortex, is quantized in multiples of 
$h/m$.
A more general, $3$-dimensional model that takes into account
the deformation of the vortex lines has been elaborated by
Fetter \cite{Fetter1}. An extension of his theory describes
vortex dynamics in extreme type {\rm II} superconductors \cite{Fetter2}.
The Hall Effect observed in type {\rm II} superconductors
  is yet another indication on the r\^ole of
 vortices in the Hall context.
 
Another derivation of the vortex Lagrangian (\ref{redlag}) is due to
Manton, who deduced it,  for large separations, from
the Landau -- Ginzburg theory \cite{Manton}.


\end{document}